# iRescU – Data for Social Good Saving Lives Bridging the Gaps in Sudden Cardiac Arrest Survival-

## Harnessing Crowd Sourcing, Geolocation and Big Data for Social Good

Nadine Levick

EMS Safety Foundation

New York City, NY, USA

nlevick@attglobal.net

**ABSTRACT**

Currently every day in the USA 1000 people die of sudden cardiac arrest (SCA) outside of hospitals or ambulances - before emergency medical help arrives - in the streets, workplaces, schools and homes of our cities, adults and children. Brain death commences in 3 minutes, and often the ambulance just can't be there in time. Citizen cardiopulmonary resuscitation (CPR) and automated external defibrillator (AED) use can save precious minutes and lives. Using public access AED's saves lives in SCA- however AEDs are used in <2% of cardiac arrests, though could save lives in 80% if available, findable, functioning, and used. The systems problem to solve is that there is no comprehensive or real time accessible database of the AED locations, and also it is not known that they are actually being positioned where they are needed. The iRescU project is designed to bridge this gap in SCA survival, by substantially augmenting the AED database. Utilizing a combination of AED crowd sourcing and geolocation integrated with existing 911 services and SCA events and projected events based on machine learning data information to help make the nearest AED accessible & available in the setting of a SCA emergency and to identify the areas of greatest need for AEDs to be positioned in the community. Helping to save lives and address preventable death with a social good approach and applied big data.

## 1.INTRODUCTION:

**BACKGROUND TO THE PROJECT PROBLEM –
SUDDEN CARDIAC ARREST AND AED GEOLOCATION**

Social good applications are often highly interdisciplinary in nature and usually require close collaboration between diverse disciplines, technical IT practitioners, subject matter experts, and social sector experts. The optimization of sudden cardiac arrest management and its systems engineering embodies this concept.

First of all, what is a SCA? It is when the heart suddenly stops pumping effectively and/or begins fibrillating or quivering and there is no effective circulation of blood and the victim rapidly loses consciousness, collapses and becomes unresponsive. Brain death commences within 3 minutes unless CPR is commenced and an AED used. However, the sad fact remains that despite major advances in technology, data management, machine learning, acute health care and emergency medical services – sudden cardiac arrest survival has hardly improved in the past 30 years across the nation. In the field of public safety, public health, community outreach and social good addressing this existing problem of survival of sudden cardiac arrest is truly a systems engineering challenge.

Of the 1000 people who die each day in the setting of an out of hospital cardiac arrest in the USA, sadly, there is a less than 2% utilization of AEDs. This is even though there is a predicted potential 80% survival rate with use of an AED in the first few minutes after a SCA, as ~80% of SCAs may have a reversible shockable rhythm with use of an AED. However, currently in the USA, for the vast majority of cardiac arrests the nearest location of a public access AED is not known. While there are estimated to be ~1 million AEDs in the USA, fewer than 10% devices exist in any database. Current average survival rates in the USA after SCA are one tenth of that 80% figure – around 8%. In the New York metro region where this pilot is being developed, cardiac arrest survival rates are amongst the lowest in the country – less than 5%. Out of hospital SCA is a major problem both nationally and in the target region for this preliminary project, with annual deaths that are over ten times those of the annual road toll.

Whilst vast resources are expended on numerous complex public safety initiatives and infrastructure such as road safety – the data management dimensions of bridging the gaps in the system of SCA survival are in need of optimization and a paradigm shift.

There are gaps in process and key information regarding: performing CPR, capture and validation of AED location, the AED functionality and accessibility and utilization of AEDs. A comprehensive, reliable, dynamic and easily accessible national AED geolocation database has yet to be created. The goal of our project is to pilot a sustainable and inexpensive geolocated AED database augmented by using new communication technologies, such as social media, QR codes, the cloud and machine learning.

Integrating the complex data of where AEDs are currently located, their functionality and accessibility, real time access to this data in a SCA emergency, both by the 911 system and the bystander public, and information on SCA event geolocation for the deployment of additional AEDs to areas of need - are the core elements of this iRescU Project.

Capturing initially regional, and ultimately global, data on AED geolocation is a challenging big data project – and the data capture process to do so is multi modal and multifaceted. Use of novel community engagement and crowdsourcing gamification in conjunction with machine learning tools to assist in identifying AED geolocation and integrating health outcomes data to facilitate predictions for optimizing AED deployment and location are both powerful and novel dimensions to this project.

## 2. INTERDISCIPLINARY ENGAGEMENT SOCIAL GOOD, THE CLOUD, HEALTH OUTCOMES DATA AND SAVING LIVES

This project is a work in progress – and heavily reliant on in depth integration and collaboration between IoT, IT technologists, machine learning practitioners, subject matter experts, emergency services, health care analytics and public health experts as well as community outreach and social media social sector expertise.

There are 3 phases to this project:

1. Building a comprehensive regional cloud based, validated and real time accessible AED database

2. Collaborating with an existing regional health care data system to identify geolocation of SCA events in that system, and also geolocation of SCA 'at risk patients' using health care analytics

3. Evaluation of the effectiveness of the development of the regional AED cloud based database compared to the current manual model, and also any measurable change in use of AEDs for SCA events in the out of hospital setting and any change in regional SCA survival rates in the two years subsequent to the deployment of the cloud based AED database.

All potential and feasible modalities are being engaged to capture existing AED geolocations and accessibility and functionality – from low tech to high tech – community crowd sourcing to machine learning tools.

• Crowd sourced geolocation data is being captured via all possible data entry tools, photos, smart phone capture, an App, and a web site. Creative use of QR codes as a data entry tool on promotional items such as t-shirts and baseball caps and key chains has been a valuable approach to date.

• Integrating and validating existing manual or electronic AED databases.

• Use of IMB Watson visual images and machine learning to seek out existing AEDs.

A collaboration with a regional electronic health data system is underway for this project – so that three objectives can be achieved

• Identification of geolocation and demographics of regional SCA events

• Analytics for population based prediction of AED location needs

• Evaluation of the SCA outcomes of deployment of this regional pilot project.

## 3. PILOT DATA AND PROOF OF CONCEPT

In our pilot work, an interdisciplinary team utilized a two-way, closed-loop, cloud-based data management system on a sequel server, configured to be populated by crowdsourced data and to upload existing, static AED databases. The preliminary work of the design and development of a cloud based AED demonstration data base has been completed, functionally linked to community generated AED geolocation data capture. Gamification of this approach has been tested in preliminary work. A number of pilot community outreach AED geolocation challenges have been conducted, which have demonstrated effective validatable crowd sourced and cloud based AED geolocation data. These pilots included engagement of SCA aware cohorts, a 5[th] grade class, and a regional university based deployment in Denver.

During the pilots 5 AED crowdsourcing contests were conducted. A number of social media strategies for outreach were used, including twitter, youtube and 2D QR codes on electronic media, business cards, postcards and also clothing and promotional items, to direct mobile devices to an AED geolocation upload form, which were distributed during contests. Mandatory fields on the upload form were street, building location, city, state and country. Photos were optional. Crowdsourced AED geolocations were uploaded from 3 continents. The range of maximum number of AED geolocations uploaded by an individual was 37- 103 respectively, during the 5 contests. The highest AED geolocators in this study were those very familiar with AEDs - parents of a child survivor, researchers in AED use, a classroom of 5th graders with AED geolocation as a project, and first-responders. Total direct cost of the pilot crowdsourcing infrastructure was $450 per contest.

Additionally this project has had recognition as being selected for the Tech Demo Pavilion at the mHealth Summit, presented at New York Tech Meet up Demo 2013, an awardee of the Inaugural American Heart Association Innovation Challenge 2014, and has been invited to the White House to participate in the Safety Data Jam and Safety Data Palooza, and most recently has been accepted to the New York Economic Development Corporation (NYEDC) SBIR 2016 Impact Program.

Partnerships with key collaborators for 2016 SBIR/STTR submission and health outcomes data linkages are currently being established, including technical and operational partners, and community engagement. Preliminary planning is also underway to identify the optimal pathway for harnessing machine learning application for IBM Watson searches of visual images for existing AED geolocations.

## 4. CONCLUSION

Out of hospital SCA is a societal public health problem in need of a disruptive solution. We should not tolerate 350,000 deaths annually in the streets, schools, playgrounds and homes of our communities, and which are ten times the road toll or the deaths from guns! This project utilizes novel data entry tools, social media, gamification and geolocation challenges, life cycle engagement, a very lean foot print and integration with existing technology tools, the cloud, health outcomes data, machine learning and societal infrastructure.

It is a creative sustainable, scalable, translatable model leveraging some existing resources, mobile and IT to help address a major public health issue.

Gamifcation and social media and mobile technology can be harnessed to support creating a global dynamic and inexpensive AED database. Social media may be a valuable tool for outreach into the broader community to increase awareness of OHCA and to leverage broader based AED geolocation data.

iRescU is one small step for mobile health and social media– one giant leap for public health and saving lives!

.

# 5. ACKNOWLEDGEMENTS

Much thanks to Prof Art Cooper for his wise guidance and encouragement for the development of this project. To Adrian Dore and Simon Ralphs of Telematicus for their talented technical skills in database development and project execution, the interdisciplinary iRescU Research Team for their most helpful input and suggestions and to the Advisory Board of the EMS Safety Foundation for their ingoing advice. Also to Evie McNee for her promotional/educational video and input, and to Medstartr for their involvement in supporting the iRescU Project and the AHA Innovation Challenge.

# 6. REFERENCES

1. American Heart Association (AHA) Nov 2011 - CPR/AED Apps: What's Out There and Are They a Reliable Source of Public Lifesaving Information? –N. Levick and the iRescU Team

2. mHealth Dec 2011 – iRescU – Bridging Social Media and Saving Lives: Logistics and Implementation of a Two-way Global Cloud Based Life Saving Platform in the Real World. N. Levick and the iRescU Team

3. Emergency Cardiac Care Update (ECCU) Symposium Sept 2012 - iRescU – Bridging New Technology, Social Media and Saving Lives: Is a Global AED Scavenger Hunt Feasible? A Brief Pilot of Use of an etagged and Social Media Approach to Disseminating Crowd Sourcing of AED Geolocation. N. Levick and the iRescU Team

4. Denver AED crowdsource geolocation gamification pilot 2012 – AED Scavenger Hunt video, https://www.youtube.com/watch?v=jgpMD4ZpGyE

5. Informs Conference June 2013 - iRescU: Bridging New Tech, Social Media & Saving Lives – A Novel System for CPR/AED Management. Development and implementation of a two way cloudbased multiplatform CPR/AED global App and Data Management System in the Real World. Nadine Levick MD, MPH EMS Safety Foundation, New York, USA, Simon Ralphs and Adrian Dore Telematicus, London UK, Stellah DeVille Stellah Inc and The iRescU Research Team

6. New York Tech Meetup Demo March 2013 – the iRescU Project, https://vimeo.com/102760497

7. Heart Rhythm Society (HRS) Conference May 2013 - The Use of Social Media and Innovative Technology to Create a Global AED Geolocation Database- The iRescU Project. N. Levick and the iRescU Team

8. American Heart Association (AHA) Innovation Challenge 2014 – iRescU Project, Second place winner. N. Levick and the iRescU Team

9. iRescU video for Medstartr – youtube 2014 by Evie McNee iRescU Scout Leader https://www.youtube.com/watch?v=JyMpFZ9v6N4

10. NYEDC SBIR Impact Program June 2016 - http://www.sbirnyc.com/ - iRescU Project selected as a participant. N. Levick and the iRescU Team